**Risk analysis beyond vulnerability and resilience – characterizing the defensibility of critical systems**

**Authors**

Prof. Vicki Bier, PhD (corresponding author)

(Department of Industrial and Systems Engineering, University of Wisconsin-Madison,

1513 University Avenue, Madison, WI, 53706 USA bier@engr.wisc.edu)

Prof. Alexander Gutfraind, PhD

Laboratory for Mathematical Analysis of Complexity and Conflicts

Epidemiology & Biostatistics, School of Public Health,

University of Illinois at Chicago, Chicago, IL, 60612 USA

agutfraind.research@gmail.com

and

The Program for Experimental & Theoretical Modeling, Division of Hepatology,

Department of Medicine, Loyola University Chicago, Maywood, IL, 60153 USA)

**Abstract**

A common problem in risk analysis is to characterize the overall security of a system of valuable assets (e.g., government buildings or communication hubs), and to suggest measures to mitigate any hazards or security threats. Currently, analysts typically rely on a combination of indices, such as resilience, robustness, redundancy, security, and vulnerability. However, these indices are not by themselves sufficient as a guide to action; for example, while it is possible to develop policies to decrease vulnerability, such policies may not always be cost-effective.

Motivated by this gap, we propose a new index, defensibility. A system is considered defensible to the extent that a modest investment can significantly reduce the damage from an attack or disruption. To compare systems whose performance is not readily commensurable (e.g., the electrical grid vs. the water-distribution network, both of which are critical, but which provide distinct types of services), we defined defensibility as a dimensionless index.

After defining defensibility quantitatively, we illustrate how the defensibility of a system depends on factors such as the defender and attacker asset valuations, the nature of the threat (whether intelligent and adaptive, or random), and the levels of attack and defense strengths and provide analytical results that support the observations arising from the above illustrations. Overall, we argue that the defensibility of a system is an important dimension to consider when evaluating potential defensive investments, and that it can be applied in a variety of different contexts.

## 1. Introduction

One of the central concerns of the fields of security studies and risk analysis is protecting a set of critical assets or a system from disruptions or attacks (E. Banks, 2005; Haimes, 2016).  The assets might be, for example, components of a critical infrastructure system, facilities, or concentrations of people; the disruption might be accidental/natural or deliberate, physical or cyber.  In the case of deliberate attacks, the defender invests resources to protect the system while the attackers select targets in order to maximize some objective function related to the damage expected from the attacks. The case of natural events is similar, except that the attack is modeled as a stationary threat rather than responding to defense investments.

It has also been observed (Bier et al., 2007a) that, in some systems, the defender may lack the ability to tangibly improve the security of a system, even with a substantial budget of resources. This may occur when many alternative valuable targets exist, and an adaptive attacker can find and switch to an unprotected target.  Indeed, certain modes of terrorism such as knife and car attacks are virtually unstoppable in the sense that the attacker can always find a weak or undefended target.

In this paper we propose to characterize such situations using the new theoretical property of *defensibility*. Namely, we call a system *defensible* if modest investment of resources can significantly improve the outcome to the defender. The value of this *defensibility* measure is that it enables the analyst to determine whether investments in defense are the best protection strategy.  If the defensibility is found to be low, the defender should instead seek alternative strategies, such as deterrence through retaliation (in the case of intentional threats) or effective emergency response.

Numerous terms have been defined (Haimes, 2016; Zakour & Gillespie, 2013) to characterize systems (Table 1).  For example, "vulnerability" has been defined as a "physical feature or operational attribute that renders an entity, asset, system, network, or geographic area open to exploitation or susceptible to a given hazard" (Beers & Risk Steering Committee, 2010), "the conditional probability of success given a threat scenario occurs" (US Department of Homeland Security, 2003, sec. 68/126), or "the degree to which a system is affected by a risk source or agent" (Aven, 2015).  Note that vulnerability is actually in general a vector-valued concept (Haimes, 2006), since a system may have different levels of vulnerability to different threats.  However, for a sufficiently well-defined threat (e.g., a particular type of attack, by an adversary with a given level of capability; or an earthquake with a given peak ground acceleration, direction and magnitude of ground motion, duration, and frequency spectrum), it may be reasonable to view vulnerability as a scalar reflecting the likelihood of damage. Similarly, Haimes et al. (1998) distinguish between resilience (the ability of a system to return to normal rapidly) (Alderson et al., 2015; Helfgott, 2018; Hollnagel et al., 2006; National Academies, 2012), robustness (ability to function despite damage) (Haimes, 2006), redundancy (spare capacity) (Ganin et al., 2016), and security (effectiveness of measures to limit access to a system).

**Table 1: Comparison of key concepts of risk analysis with defensibility**

| Concept | Motivating Question(s) |
|---|---|
| Defensibility | To what extent would strengthening the system reduce the attack damage? |
| Robustness | How well can the system absorb damage? |
| Resilience | How well could the system recover from damage? |
| Security | How likely is it that malicious attackers would be interdicted? |
| Vulnerability | How likely is it that a disruptive event or attack would cause damage? |

In this paper, we propose a new characteristic—defensibility. We see defensibility as being in some sense related to changes in expected damage with defensive investment. A system can have high expected damage, but be easy to defend, or alternatively can be at low risk of damage but difficult to defend. Based on this, we argue that the defensibility of a system is an important dimension to consider in security analysis.  The concept of defensibility as defined below is simple but novel. Defensibility has been previously used only as a *qualitative* term to discuss the protection of territory in military contexts (Ljung et al., 2012) and metaphorically in business, law and management strategy. Some authors have looked at risk reduction as a function of budget, but in the context of a single system (Jonkman et al., 2003), whereas here we propose to quantify defensibility for the purposes of comparing systems.

In light of defensibility, identifying system vulnerabilities may be of relatively little value if the system is not highly defensible (although of course knowing the vulnerabilities is necessary to calculate defensibility).  For example, a highly vulnerable system may be defensible against some vulnerabilities but not others, or may not be readily defensible at all.  To illustrate this latter point, Salmeron et al. (2004) computationally identify the attack strategies that would cause the maximal disruption to electricity systems, and propose protecting the system against those attack strategies, arguing that ''By considering the largest possible disruptions, our proposed plan will be appropriately conservative.'' However, one of us observed that hardening even a significant percentage of an electricity system may not dramatically diminish the damage (i.e., load shed) as the result of an intelligent attack (Bier et al., 2007a). Moreover, hardening of individual assets may sometimes be less effective than overarching forms of protection such as border security (Haphuriwat & Bier, 2011). Thus, it is not clear that identifying the most damaging attack strategies will always be a helpful guide to system hardening, and other strategies may need to be considered.

The organization of the paper is as follows. The next section defines defensibility. Section 3 applies defensibility to the special case of discrete assets, where the value of the system is simply the sum of the asset values. Section 4 explores in depth the properties of defensibility using the examples of two systems and a variety of attackers. Section 5 analytically proves a variety of properties of defensibility in the discrete asset case.  Finally, the paper discusses the results and concludes.

## 2. General Formulation

To define defensibility, we require an attacker, a defender and a system. We presume that the attacker has a certain capability or severity that could result in damage to the system. In our terminology, we use the word "attack" or "attacker" to refer to deliberate and non-deliberate disruptions, hazards or threats, including natural and accidental events.  The defender has a certain capability to prevent damage and derives certain performance or value from the system. The severity of the damage generally depends on the defender and the characteristics of the system. Note that we do not limit our analysis of deliberate attacks to the case of zero-sum games; in fact, Section 4.4 below explicitly discusses how the results change when the attacker and the defender have radically different objective functions.

We define defensibility as *the ability of the defender to reduce the damage to the system* below some given initial level (corresponding to any preexisting defenses) using a given level of defense effort. By *damage* we mean the difference between the initial and final system after an attack or adverse event. The final value is also termed "residual value" and is measured in units appropriate for the system in question such as dollars for a set of economic activities, kilowatts for electrical transmission and even lives when protecting human lives. The residual value increases with defense investments (measured in resource units such as dollars), and so defensibility is concerned with how much increasing the defense effort would change the residual value of the system (Fig. 1). The outcome of the attack need not be deterministic, and may be subject to stochastic variation; similarly, the defense investment might be variable or even a game-theoretic mixed strategy (Lempert et al., 2016). Such stochastic settings can be addressed by treating the residual value of the system as an expectation.  For calculations of expected damage in various practical cases, see Jonkman et al. (2003). Note, however, that the residual value of a system in a stochastic setting need not be limited to an expectation, which may not be a good metric for low-probability, high-consequence risks (Sarin & Weber, 1993).  For example, prospect theory (Kahneman & Tversky, 1979; Sunstein, 2003) suggests that public perceptions of terrorism risk are relatively insensitive to probabilities, so defensibility might be defined to focus primarily on reducing the maximum possible damage, without regard to likelihood.

From this perspective, defense effort could improve the residual value of the system in question by reducing the damage from attacks.  In principle, that could be accomplished by multiple different means – all of which can in principle be encompassed within our framework:

1. Improving security (increasing the likelihood of interdicting an attack);
2. Reducing vulnerability (reducing the success likelihood of a given attack type);
3. Improving robustness (reducing damage given a successful attack);
4. Improving resilience (reducing the duration of damage); or even
5. Increasing deterrence (reducing the likelihood that an attack will be attempted).

The concept of defensibility complements the effort to identify critical assets to be protected (see for example Ayyub et al., 2007; Banks & Hengartner, 2014; Izuakor & White, 2017), particularly when those assets belong to more complex systems (Apostolakis & Lemon, 2005; McGill et al., 2007).  In line with that approach, we define defensibility for systems, not individual assets or targets.  Hypothetically, one might think of an asset as highly defensible if the likelihood or magnitude of damage to that asset can be reduced dramatically with only a modest investment—but that is a special case of what we define here.  Rather, we are interested in quantifying the defensibility of an entire system with multiple assets. While investment may be implemented at the asset level (e.g., a security barrier), the benefit of that investment can be viewed as increasing the residual value of the system as a whole. Thus, the concept of defensibility that we propose here is an emergent property that can relate in complex ways to the individual assets of a system.

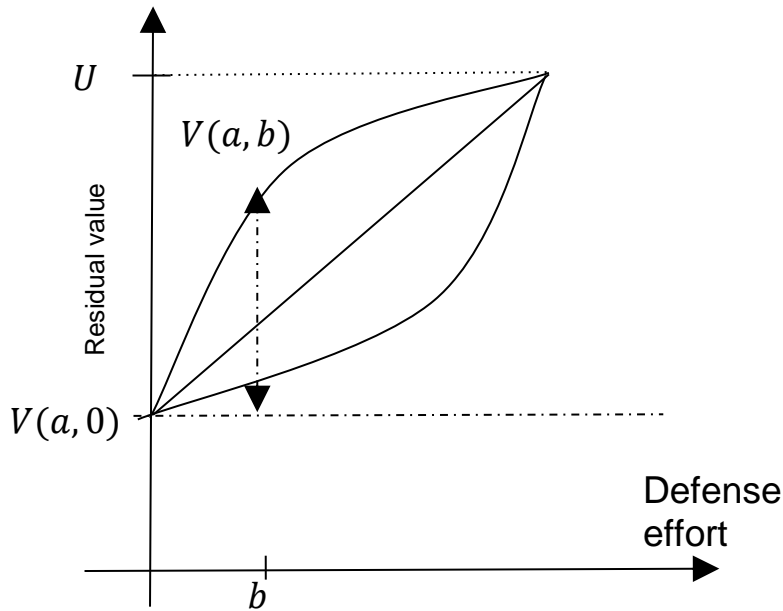

**Fig. 1. Hypothetical residual value curves for three similar systems having differing levels of defensibility.** $V(a,0)$ and $V(a,b)$ are the residual values of the system after an attack effort $a$ in the cases of zero defense effort and defense effort $b$, respectively. For example, the upper concave curve represents a highly defensible system, where a small defense effort results in a large increase in the residual value of the system. Defensibility is proportional to the difference between the curve for residual value and the horizontal line corresponding to $V(a,0)$, as indicated by the double arrow.

## 2.1 Quantification of defensibility

Let $V$ be a function expressing the value of the system to the defender. $V$ may be expressed in system-specific units, such as the number of military bases, megawatts of electrical-generation capacity, or ton-miles of cargo throughput. For a given attack, with a given likelihood and severity of damage, the resulting value of the system would be degraded (e.g., reflecting loss of services), but if the system was defended, the value after an attack is expected to more closely approach the value before any attack.

Numerically, let $V(a,b)$ be the residual value of the system to the defender when the attacker has strength $a$ and the defender makes investment $b$. If the defender makes no defense investments, the residual value is $V(a,0)$. If neither player threatens the system, the system has a nominal or "initial" value $U$, where $U = V(0,0)$. From here we define the defensibility of the system by the following function:

$$D(a,b) = \frac{V(a,b) - V(a,0)}{U} \qquad (1)$$

(Strictly speaking, defensibility is also a function of the initial level of defensive investment $b_0$, but we suppress that in the notation because $b_0$ is essentially a known constant from the perspective of our model.) Defensibility is normalized to allow comparison of systems where value is expressed in different natural units, rather than requiring value to be converted to common units such as dollars or utilities. The normalizing constant is chosen to be the full value of the system with no attack or defense.

This function represents the increase in residual value due to defensive investment (as a fraction of the initial system value $U$). The defensive investment $b$ may be discrete or continuous. For example, consider two systems, embassies and military bases, both threatened by hostile militias. Comparing the improvement in the two systems for a large value of $b$ may reflect programmatic decisions (such as whether to invest in protecting embassies or military bases), while the comparison for a small value of $b$ may help to support decisions about incremental increases in investment (e.g., whether it is worthwhile to improve the protection of one additional embassy). The denominator ensures that the expression is dimensionless and bounded in $[0,1]$, allowing incommensurable systems to be compared to each other using a single measure. Thus, the analyst need not specify quantitative differences in importance between systems that are believed to be comparable in criticality. Since the denominator $U$ is a constant for each system, the measure essentially tracks the effects of attack effort (*a*) and defense effort (*b*) on the numerator of the function. Here, cost is taken into account by the fact that the decision maker must allocate a fixed defensive effort or budget *b* to one or another system; defensibility then reflects the impact that allocation would have.

This definition of Eqn. 1 could be applied to a variety of systems and settings. In the next section we study a restricted setting with discrete assets (all with the same protection cost) and binary protection decisions. In this case, the smallest possible investment is $b = 1$; i.e., protection of one asset. More generally, one could consider continuous investment decisions (in which case we might be concerned about infinitesimal increases in $b$), or cases where the assets have different protection costs. For an example where defensibility could be considered as a function of continuous monetary investments, see Levitin (2009), where beyond a certain level of defensive investment, the system defensibility becomes zero (in the sense that no further investments can be justified).

Regarding the value function $V(a, b)$, there are several important analytical cases. First, there is the case where the system value is simply the sum of the values of the individual assets, which we analyze in the next two sections. A more complex non-additive case is that of a completely connected network. For example, Metcalfe's law (reviewed in Zhang et al. 2015) postulates that the value of a network will be a quadratic function of the number of nodes, due to synergies associated with being part of a network. In the general case of complex systems (Page & Miller, 2007; Przemieniecki, 2000), the quantity $V(a, b)$ may not be an analytical expression, and may need to be computed from numerical simulations of the system and possible attacks; e.g., using system-dynamics models or discrete-event simulations (Ganin et al., 2016; Gutfraind, 2010).

In the next two sections we focus on cases where the defender can perfectly protect any defended asset, and simply deflects any optimizing attackers to other assets. However, in more complex networked systems, the level of defensibility could be affected by how the damaged components change system functionality. For example, increasing traffic density on undamaged components could lead to congestion and increased travel times, while increasing the electricity flow on undamaged components could lead to cascading failures and blackouts. In such cases, the effect of protecting a particular asset may depend not only on its valuation (e.g., its load-carrying capacity), but also on the network topology and the position of that asset in the network. The definition of defensibility given above still applies in such cases.

## 2.2 Defensibility as an optimization criterion

### 2.2.1 The system selection problem

One of the most important applications of defensibility is for evaluating defensive investments across systems. Frequently, policy makers face the following scenario:

- A limited defensive budget $b$ needs to be allocated among several systems, and it may be desirable to allocate the entire investment to a single system (e.g., due to of commitment costs or administrative reasons)
- The systems' values are not directly or easily comparable to each other (for example, the electrical grid and the water-distribution network are both critical, but their values are not directly comparable, since they provide distinct services), but policy makers would like a metric that allows them to compare systems that are all believed to be comparably critical. The threat is unlikely to switch from one system to another in response to the observed defensive investment, at least within the planning horizon (for example, terrorists that attack civilian targets may lack the capability or intent to attack hardened military installations). If that assumption is violated, then we are in the arena of systems-of-systems, and the defender may wish to calculate the defensibility of the entire system of systems at a higher level of analysis.

Defender optimization problem. This problem could be formulated as a constrained optimization problem, as follows. The objective is to maximize the total residual value of all critical systems, where the values of all systems are roughly equal, subject to the constraint that a single system must receive the entire defensive budget $b$. After calculating the defensibility of all systems, the optimal policy is to invest in the system with the highest defensibility.

2.2.2 Defensibility in system design

Another important application of defensibility is in the context of system design. When a system such an infrastructure network is being designed, the planner must consider a variety of metrics in the design process – chiefly performance and cost. The existence of threats to the system often adds system vulnerability to the design considerations. However, cost or other pragmatic considerations may make it infeasible to design a system to achieve low levels of vulnerability. In that case, the design of the system can aim to achieve high defensibility – the ability to upgrade the system at a modest cost in order to reduce its vulnerability in the future. Designing for defensibility is analogous to buying a real option (Trigeorgis & Tsekrekos, 2018) for lower vulnerability in the future.

High defensibility might be less onerous to achieve than low vulnerability, and so attractive to designers of a variety of systems. For example, warships may be designed to support future defensive upgrades in case additional defenses become essential at a later time (e.g., if adversaries acquire more effective anti-ship weapons). Similarly, desktop computers can be easily upgraded when new vulnerabilities are identified by updating the antimalware software. By contrast, one concern that has been raised regarding the Internet of Things (the growing network of smart and interconnected devices) is that it has not been designed with defensibility in mind, since many of these devices cannot be easily upgraded (Jing et al., 2014).

### 3. Defensibility for Systems with Discrete Assets

In this section we apply our ideas to an important special case where the system consists of discrete assets and the value of the system is just the sum of the values of the assets. Defense is assumed to be binary, so that a defended asset maintains its full value, while an attacked asset is assumed to be fully destroyed (value of 0). These assumptions (additive system valuation, and perfect binary defense) clearly represent a simplified case, for illustrative purposes. Note also that readers more interested in applications can proceed directly to Section 4 below, which illustrates the idea of defensibility through numerical examples with realistic data.

We consider in detail both an optimal and a uniform random attacker; of course, our model also admits intermediate attack strategies (such as an attacker that prefers high-valued assets, but does not observe which assets have been defended, or an attacker that values all assets equally, but observes which assets have been defended and avoids attacking them). While there are a variety of possible game-theoretic models of security (Bier & Azaiez, 2009), here we focus on a sequential two-stage game with perfect information; for a more general discussion of game theory, see Myerson (2013). In particular, the defender selects which asset(s) to defend, while the attacker observes the defender's choice(s) and chooses which asset(s) to strike in order to maximize the total damage. In the discrete setting considered here, this means that the attacker destroys the most valuable $a$ of the undefended assets. This is a common but conservative assumption, since in practice, an attacker might attack a less valuable asset (e.g., due to insufficient information about the defender's asset values). Alternatively, one could frame the problem in terms of adversarial risk analysis (Banks et al., 2016), which avoids the assumption that the defender has perfect information about the attacker's asset values, and simply allows the defender to choose the best possible defense in light of the defender's beliefs about the attacker's asset values. Moreover, the random case can be used to represent disruption due to accidents or natural disasters that are equally likely to affect any asset, as well as disruption due to an attacker who is indifferent or uninformed.

### 3.1 Notation

$n$        Number of assets in the system

$a$        Number of assets attacked or disrupted

$b$        Number of assets defended

$v_i$       Value of asset $i$ to the defender, $v_1 \geq v_2 \geq \cdots \geq v_n$

$w_i$       Value of asset $i$ to the attacker

$A_i$       Probability that asset $i$ will be attacked or experiences disruption

(Note that according to the subjectivist definition of probability, this can be taken to be the defender's subjective probability of attack or disruption, since in reality an objective or frequentist probability is likely to be unavailable)

$B_i$       Indicator variable equal to 1 if asset $i$ is defended, and 0 otherwise

$s_i$       Probability that asset $i$ survives

$v_i(a, b)$ Expected value of asset $i$ following an attack or disruption of the system

$V(a, b)$ Expected remaining total value of the system after an attack or disruption

$U$        Initial value of the system, before any attack or defense, $U = V(0,0)$

### 3.2 Reformulation of defensibility for discrete assets

It follows that

$$s_i = (1 - A_i) + A_i B_i \tag{2}$$

The expected value of asset $i$ is then

$$v_i(a, b) = v_i s_i(a, b) \tag{3}$$

The expected remaining total value of the system is just the sum over the assets

$$V(a, b) = \sum_i v_i(a, b) = \sum_i v_i s_i(a, b) \tag{4}$$

Finally, defensibility becomes:

$$D(a, b) = \frac{\sum_i v_i s_i(a, b + b_0) - \sum_i v_i s_i(a, b_0)}{U} \tag{5}$$

### 3.3 Types of attackers, defenders and systems

For deterministic optimal attackers, we consider two cases:

(1) A same-value attacker assigns the same value to each asset as the defender, resulting in a zero-sum game.

(2) A different-value attacker receives a benefit $w_i$ for destroying asset $i$, where in general $w_i \neq v_i$. This type of attacker need not satisfy $w_1 \geq w_2 \geq \cdots \geq w_n$, and the game is not zero-sum.

A **stochastic attack** refers to any disruption (deliberate or natural) that strikes targets non-deterministically. A **stationary attacker** is a special case of a stochastic attacker whose probability distribution of attacks does not depend on the defender's choices. A **uniform random attacker** is a stochastic stationary attacker that strikes targets with equal probability; i.e., completely at random. (Note that in the case of a deterministic attacker, whose behavior can be completely predicted from knowledge of the attack effort $a$ and the defense effort $b$, the attack probability $A_i$ will simply be an indicator variable equal to 1 if asset $i$ will be attacked and 0 otherwise—at least if we exclude the theoretical case where multiple assets have the same value to the attacker, leading to random attack strategies even in the case of perfect information about the attacker.)

We will see shortly that the distribution of asset values has a strong effect on the defensibility of a system. The most important limiting case is that of convex decreasing values, where the asset values exhibit decreasing differences in value; i.e., $v_i - v_{i+1} \geq v_{i+1} - v_{i+2}$ for all $i = 1 \ldots n - 2$. The other limiting case is of concave decreasing values. Asset values with decreasing differences will exhibit positive skewness (MacGillivray, 1986). This reflects the typical situation with a few high-valued assets and a large number of low-valued assets. (Of course, not all positively skewed sets of asset values will exhibit decreasing differences, since skewness is a global rather than a local property.) By comparison, asset values with increasing differences will have negative skewness, reflecting a situation with a large number of near-optimal targets, and only a few low-valued targets.

Yet another situation is when the attacker can choose between attacking a given system of interest to the defender and attacking some other system. In this case, the attacker can be modeled as having a nonzero "opportunity cost" of attack, such that the system of interest to the defender will no longer be targeted if the maximum possible attack damage achievable after defense is less than the attack damage resulting from an attack on the other system. Here, we are again in the case of a system-of-systems.

## 4. Defensibility of Infrastructure Systems

### 4.1 Data sets

We apply the above model to two representative data sets exhibiting positive skewness, as well as to a synthetic data set with negative skewness (for illustrative purposes):

**Property Losses**. We use data from Willis (Willis, 2007; Willis, Morral, Kelly, & Medby, 2005), which provides estimates of the expected annual property losses that would result from attacks on different urban areas in the United States. For simplicity, we restrict our attention primarily to the 10 urban areas of the United States that are estimated to have the highest expected annual terrorism losses in Willis; see Table 2 below. Note also that the estimates in from Willis are expected values taking into account multiple possible attack types and targets within any one urban area. However, for purposes of illustrating defensibility, we treat them as if they represent attacks on discrete assets (e.g., a single signature building in each urban area). Examination of the Property Losses data set finds that it generally exhibits decreasing differences (i.e. convexity), except for the extremely small differences between Washington and LA, and between Philadelphia and Boston.

**Air Departures.** We also consider data on the air transportation system. In this data set, the value of an airport is characterized based on the number of departures from June 2015 to June 2016 (US Bureau of Transportation Statistics, 2016) (Table 2). In particular, in each of the 10 urban areas with the highest expected annual terrorism losses (according to Willis), we consider the airport with the largest number of air departures.

**Negative Skew.** Finally, to illustrate defense of systems with a large number of similarly-valued targets, we constructed a data set exhibiting negative skewness, with asset values of 719, 712, 705, 694, 676, 655, 621, 585, 528, and 413. Note that these values show increasing differences; i.e., the asset values are concave and decreasing.

The three data sets can be usefully characterized based on the skewness of the asset valuations. In particular, Property Losses, Air Departures, and Negative Skew have skewness values of 2.8, 2.4, and -1.4, respectively. This has significant implications for the defensibility of those systems, as shown below.

## 4.2. Same-value deterministic optimal attackers

We begin for clarity of illustration with a same-value attacker. Consider the example below (Fig. 2), where the vertical axis represents residual value of the system (as a percentage of the total value $U$). The blue lines correspond to the case where residual value is measured in terms of property damage, while the difference between the dotted blue line and the solid blue line shows the defensibility of the system, $D(1, b)$, against an attacker of strength $a = 1$ for any given value of $b$. In this case, while even a single attack can do quite a bit of damage, the extreme skewness of the data set (2.8) is associated with high defensibility, even for small values of the defense effort $b$. In fact, the average defensibility over all values of $b = 1 \dots 10$ is more than 53%.

To illustrate how these results are computed, the total value of all expected property losses for the 10 cities is $719 million. If no assets are defended, the attacker is assumed to attack New York and cause $413 million in damage. Thus, the residual value is $719 - $413 = $306 million, or about 42.6% of the total value at risk. Similarly, when one asset is defended, the defender is assumed to protect New York. Since the attacker can observe the defenses, an attack is launched against the second-best target, Chicago, causing only $115 million in damage. At that point, the residual value of the system would be $719 - $115 = $604 million, or 84% of the total system value. The same logic applies as larger numbers of targets are protected.

Fig. 2 also shows that both residual value and defensibility are monotonically increasing with $b$. In fact, this is provable in general (see Prop's. 2 and 4 in section 5).

On the same figure, the red curve (for air departures, with a skewness of 2.4) shows an average defensibility over $b = 1 \dots 10$ of 11%. This value is smaller than for the property values because the data set for air departures does not have a single extremely high-valued target, thereby reducing the benefit that can be achieved by a modest defense effort. (For example, property losses for an attack on New York are nearly 3.6 times as large as for the next highest urban area, while air departures from O'Hare are only 1.5 times those from LAX, the next most significant asset in the airports data set.)

**Table 2: Data sets for property losses and air departures**

| Urban Area | Expected Property Losses ($millions) | Air Departures (thousands/year), and airport code | Population (millions) |
|---|---|---|---|
| New York | 413 | 166 (LGA) | 9.3 |
| Chicago | 115 | 375 (ORD) | 8.3 |
| San Francisco | 57 | 172 (SFO) | 1.7 |
| Washington, DC-MD-VA-WV | 36 | 140 (DCA) | 4.9 |
| Los Angeles-Long Beach | 34 | 248 (LAX) | 9.5 |
| Philadelphia, PA-NJ | 21 | 171 (PHL) | 5.1 |
| Boston, MA-NH | 18 | 156 (BOS) | 3.4 |
| Houston | 11 | 183 (IAH) | 4.2 |
| Newark | 7.3 | 158 (EWR) | 2.0 |
| Seattle-Bellevue-Everett | 6.7 | 174 (SEA) | 2.4 |
| **Total** | **719** | **1,943** | **50.9** |

Expanding our simple illustration, we now consider the defensibility for multiple levels of attack effort $a$ ($a = 1, 2, 4, 10$) (Fig. 3). Defense is more effective for large $a$, but with diminishing marginal returns: the average defensibility over $b = 1 \ldots 10$ is 53% for $a = 1$, 76% for $a = 4$, and only 88% for $a = 10$. This is because while total damage of course grows in attacker effort $a$, for large $a$ the additional attack effort is focused on relatively low-valued targets. This suggests that defensibility is more sensitive to the level of defensive effort, $b$, than to the number of targets attacked, $a$, when the defensive effort is small; again, we have been able to show that this is indeed the case (see Prop. 5 in section 5).

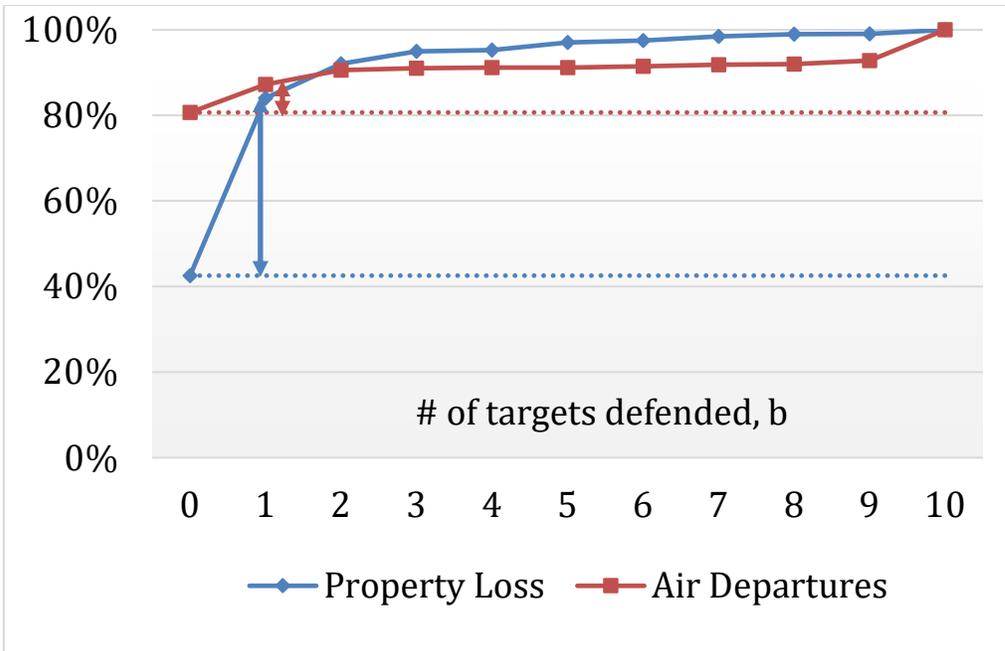

**Fig. 2. Residual value and defensibility for two systems under varying defense effort for $a = 1$.** Property Loss (blue) exhibits a large increase in residual value when defense investment is increased between $b = 0$ and $b = 1$, as compared to Air Departures (red), indicating that property is more defensible.

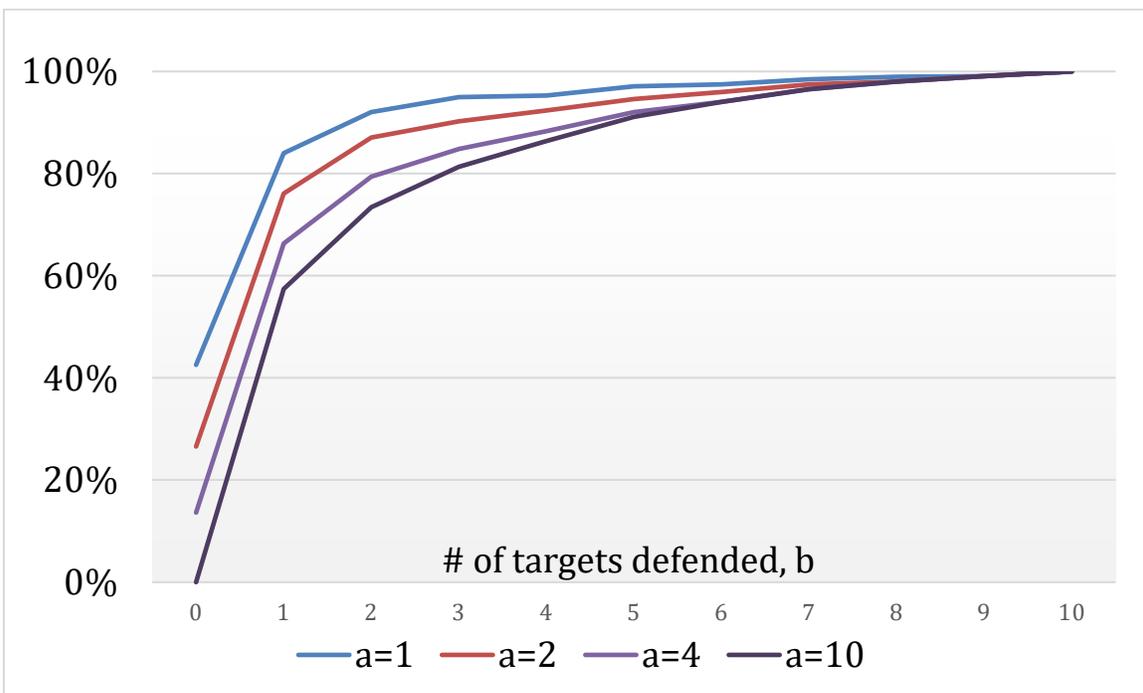

**Fig. 3. Residual value under varying levels of attack and defense effort, for an optimal attack on the Property Losses data set.**

### 4.3 Stochastic attacks

As mentioned earlier, the concept of defensibility can be applied not only to intentional attacks, but also to random (non-strategic) attacks, accidents, or acts of nature. Fig. 4 below compares the residual values for optimal and uniform random attacks for the highly skewed data set on expected property losses, and finds that the system is actually significantly more defensible against an intentional attack (i.e., an attack targeting the undefended city with the maximum expected property loss) than it would be against a threat that attacked all ten cities randomly (an average of 53% over $b = 1 \ldots 10$ for intentional attacks, vs. less than 9% for random attacks).

In particular, we have been able to show that defensibility will be higher against optimal than uniform random attackers when the number of assets is sufficiently large (where "large" depends on the attack effort $a$ and how quickly the asset values decline; see Prop. 9 in section 5). To understand why this is true, note that when the number of possible targets is large as compared to $b$, random attacks would be more likely to occur at undefended assets, making defense relatively ineffectual. By contrast, intentional attacks will tend to focus on highly valuable targets, making attacks more damaging but increasing the benefit of investing in defense – a finding important to counter-terrorism.

We now return to the question of how attack strength affects defensibility for a stochastic attacker. When attacks are random, the residual value after an attack depends more strongly on the number of attacks $a$ than in the deterministic case; compare Fig. 5 for the random case with Fig. 3 for the deterministic case. For the uniform random case, defensibility averaged over $b = 1 \ldots 10$ is less than 9% for $a = 1$, but is nearly 88% for $a = 10$; for the deterministic case the change is smaller (53% vs. 88%). (See also Prop. 10 in section 5 below.)

### 4.4 Differing attacker and defender valuations

In the general case, the attacker and defender may assign different valuations to the various assets or targets. The mismatch may be the result, for example, of some targets being of symbolic value to the attacker (even if their value to the defender is relatively low), or simply of the attacker lacking accurate information about the value of targets to the defender. For example, the attacker may wish to attack an airport, but may not know which airports have the most departures, or may prefer to target airports in cities with large populations (perhaps because of a belief that attacks on airports in high-population cities will generate more publicity). In such cases, the impact on the defender could still be measured in air departures, even if the attacker is not targeting the busiest airport.

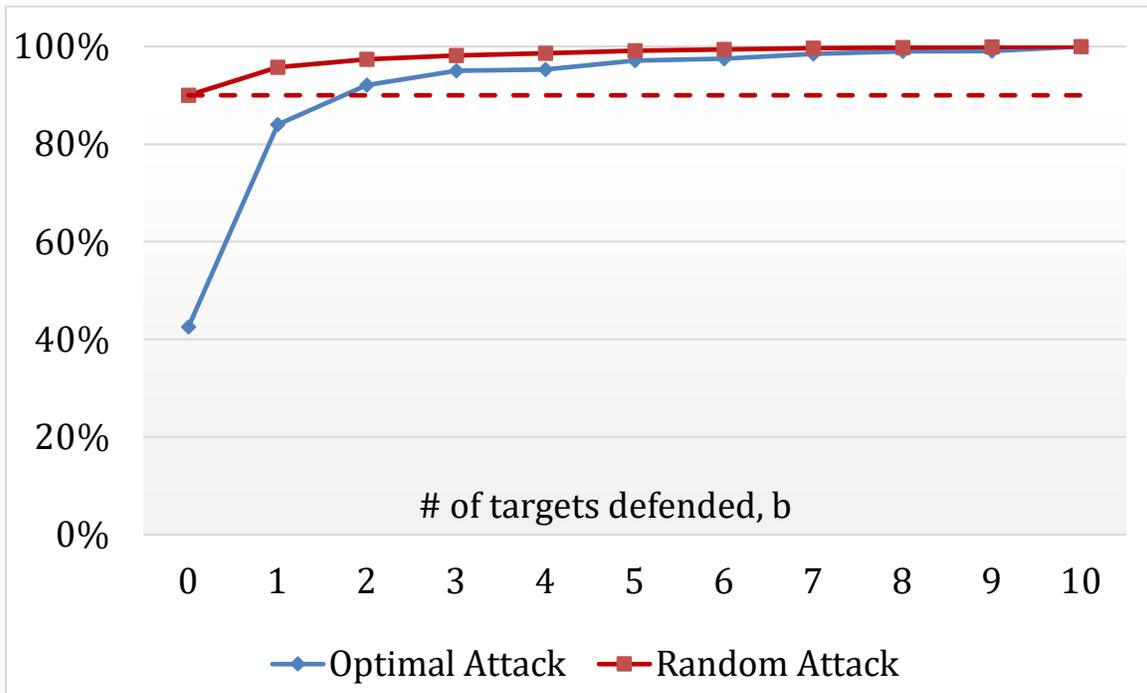

**Fig. 4. Residual values after an optimal and a uniform random attack on the Property Losses data set.** Defensibility is clearly lower against a uniform random attacker than against an optimal attacker in this case.

In our model, we find that a mismatch where the attacker chooses which airport to target based on population rather than air departures makes defense less efficient; i.e., a higher number of targets *b* must be defended to yield equivalent residual value, if the defender is unaware of the attacker's objective function (Prop. 3 in section 5). The situation is illustrated in the graph below (Fig. 6). In this case, when defending against a single attack ($a = 1$, the uppermost blue curve), the residual value does not improve at all until the defender has protected the top *two* targets (instead of getting benefit from protecting a single target), since defense of the first city (Chicago, with the largest number of air departures) was misallocated, given the goals of the attacker. Moreover, the average defensibility for $a = 1$ (over $b = 1 \dots 10$) is only about 5% (compared to 11% if both the attacker chooses which asset t target based on the defender values).

Since the calculations are more complicated when attackers and defenders have different valuations, we provide an illustration of the process. Referring back to Table 2 above, when no assets are defended and the attacker can choose only one target, the attacker is assumed to target the LAX airport, since Los Angeles is the most populous of the ten cities, and therefore causes disruption of the 248,000 air departures per year at LAX. Thus, the residual value is 1,943,000 - 248,000 = 1,695,000 remaining air departures for the duration of the disruption, or about 87.2% of the total (corresponding to the blue line in Fig. 6 below). However, when one asset is defended, the defender is assumed to mistakenly protect the Chicago airport, since it has the highest number of air departures. Therefore, no benefit is obtained from the defense, and the residual value is still 87.2%. Only when two assets are defended does the defender begin to protect LAX in addition to Chicago, since LAX has the second highest number of air departures. At that point, the attacker moves on to target La Guardia, since New York has the second highest population, and causes disruption of the 166,000 air departures per year at La Guardia. At that point, the residual value of the system would be 1,943,000 - 166,000 = 1,777,000 air departures, or 91.5% of the maximum possible system value.

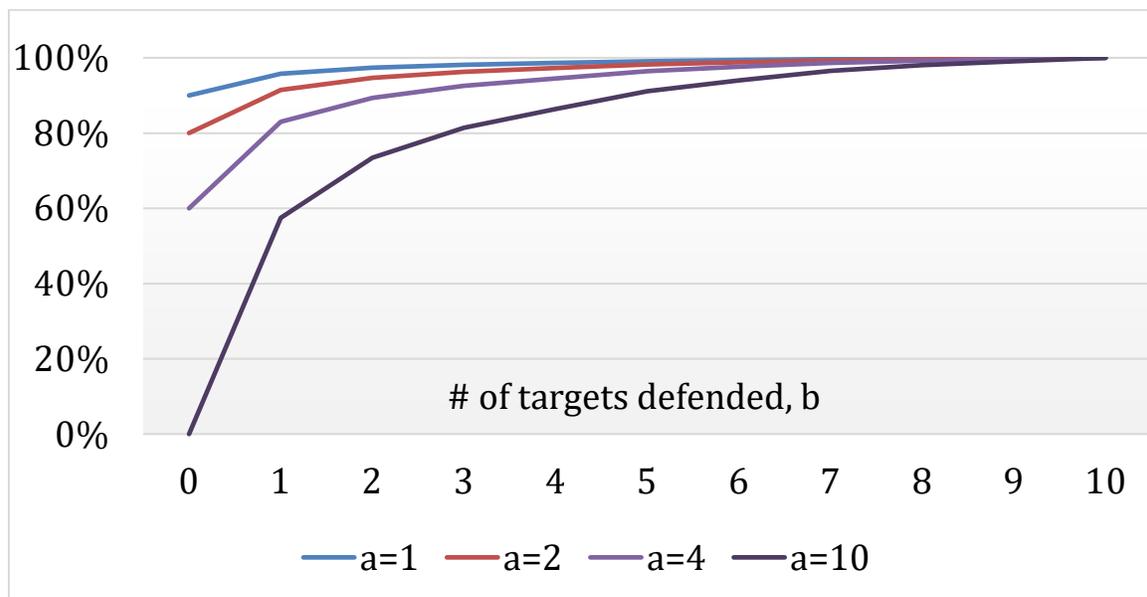

**Fig. 5. Residual value under varying levels of attack and defense effort, for a uniform random attack on the Property Losses data set.** As in the case of an optimal attack, the residual value is monotonically increasing with $b$ and decreasing with $a$ (see Prop. 7 for details), and defensibility is increasing in both $a$ and $b$ (Prop. 8).

A defender could of course attempt to improve defensibility by protecting the assets most likely to be attacked, instead of the most valuable assets. However, this strategy has some obvious pitfalls, including the fact that attacker valuations are typically not known. Moreover, when the attacker and defender valuations differ, protecting the asset(s) that are most attractive to the attacker can in some cases cause the attacker to shift towards assets that are less attractive to the attacker, but more damaging to the defender (Bier et al., 2007b).

## 4.5. Negative skew

As discussed earlier, we also apply the concept of defensibility to a system with a large number of comparably high-valued targets. Clearly, for this data set defensibility against an optimal attacker is low until large numbers of components $b$ have been defended (blue line in Fig. 7), and average defensibility over $b = 1 \dots 10$ is only about 2.5%.

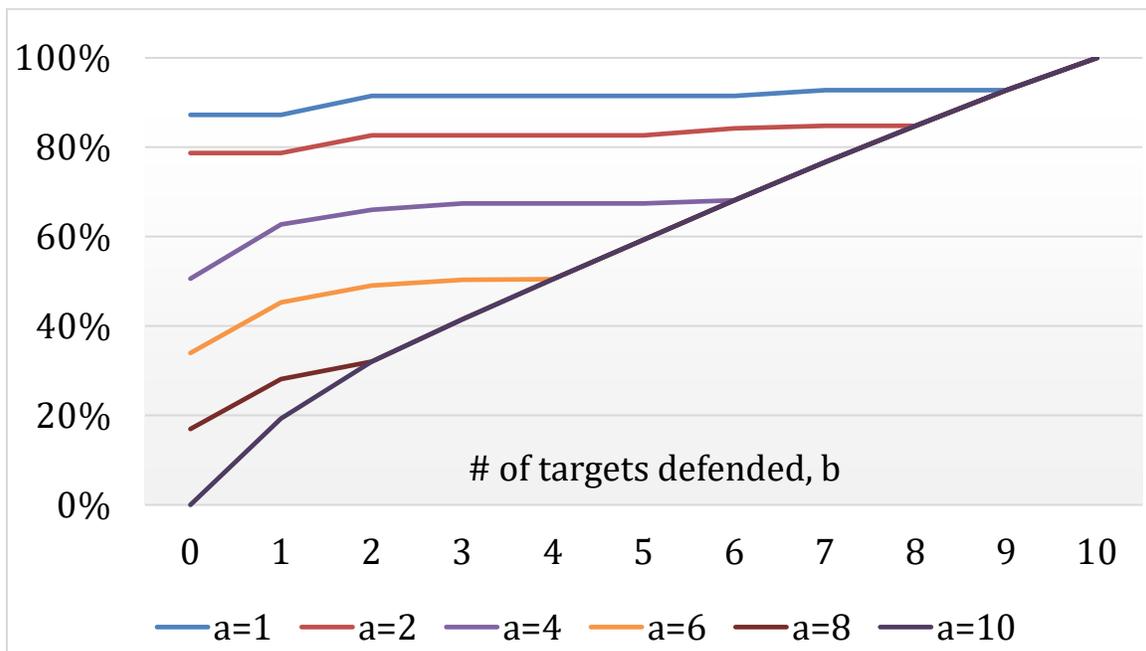

**Fig. 6. Residual value for mismatched attacker and defender target values.** When a mismatch occurs, the residual value is generally higher than in the case with no mismatch, but the defensibility is generally lower (and is often constant in $b$, or even zero).

By contrast, defensibility is actually greater against random (non-strategic) attacks than against an intentional (optimal) attack—only 2.5% (averaged over $b = 1 \dots 10$) for intentional attacks, versus 5.9% for random attacks; see Prop. 9. This is because with negative skewness, there are many near-optimal targets with almost the same values, so against an optimizing attacker, defense merely deflects the attacker to a similar undefended target. The situation is different with the random threat because it does not focus only on high-value targets and will not be displaced from attacking defended targets. Thus, in the negative skew case, defending against the random threat gives small but steady improvements in the residual value of the system.

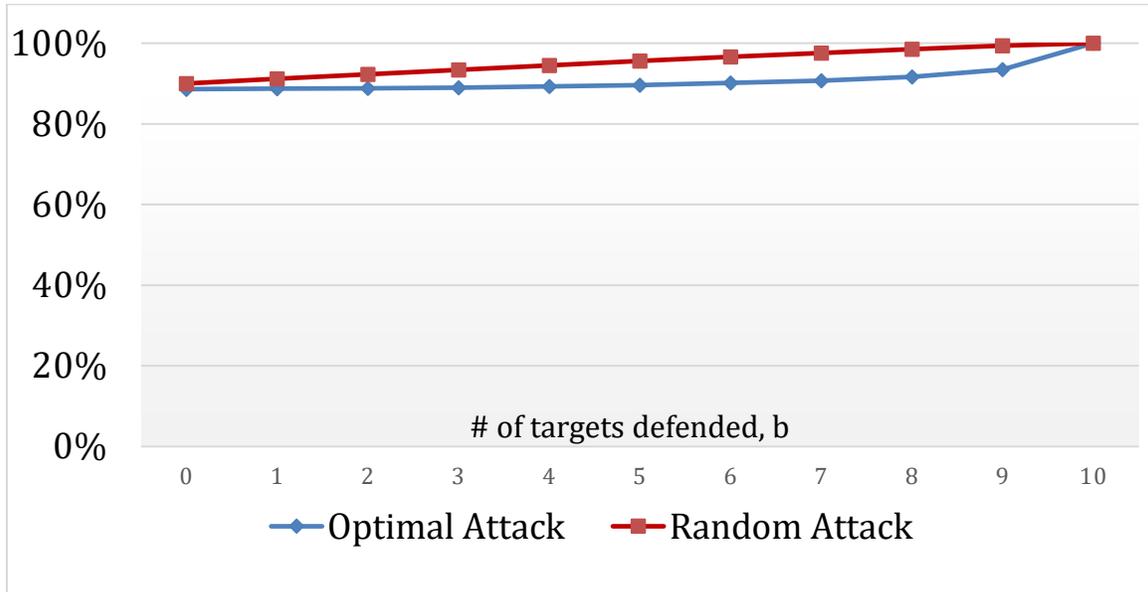

**Fig. 7. Residual values after an optimal and a uniform random attack on the Negative Skew data set.** Unlike the previous examples, which all involved data sets with positive skewness, defensibility against a random attacker in this case is actually greater than against an optimizing attacker.

## 5. Analytical Results

This section proves analytical results about residual value and defensibility. The proofs are generally obtained directly from the definition. Unless noted otherwise, we work in the setting where the system value is an additive function of the asset values, and the attacker and defender assign the same values to the assets. The assumption of equal valuations is relaxed in Property 3 below, which explores the implications of this assumption on a system's defensibility.

In many settings, we will see that the optimal defensive strategy is "reflexive," in the sense that the defender considers the values of her assets, and protects the $b$ highest-ranked (i.e., highest value) assets. This is perhaps also the easiest case to understand. In the alternative strategy of "predictive" defense, the defender defends assets by considering which of them are most likely to be attacked. However, predictive defense may be difficult to implement effectively, since it requires intelligence about the attacker's targeting. Fortunately, reflexive defense, which does not require this information, is provably optimal in certain circumstances.

Note that in what follows, we set the initial defense $b_0$ to 0 without loss of generality. This is because under our assumption of perfect defense, assets that have already been defended will never be attacked by an optimizing attacker, and will never be damaged by a uniform random attacker. Thus, we can ignore assets that have already been defended.

## 5.1 Deterministic attackers
**Property 1: Optimality of reflexive defense**

The reflexive defense strategy is optimal when (1) the attacker is optimizing, and (2) the attacker has the same values for all assets as the defender (or at least ranks the assets in the same order).

**Proof:** The optimal attacker will always strike at the $a$ highest-ranked undefended assets, where ranking refers to their values: $v_1 \geq v_2 \geq \cdots v_n$. Consequently, to maximize the residual value $V(a, b)$, the optimal defense would minimize the total value of the assets targeted by the attacker. It follows that defending the $b$ highest ranked (i.e., highest-value) assets is an optimal strategy. QED.

Note that neither of the conditions stated in Property 1 is sufficient by itself; alternatives to either one may lead the reflexive strategy to be non-optimal. (1) A non-optimizing attacker could respond unpredictably to a defense, and therefore could even adopt a more damaging strategy when the defender chooses the reflexive defense. (2) Consider the case where $a = 1, b = 1$ and the defender asset values are 10,5,1, while the attacker values for those assets are 1,10,5, respectively. In that case, the reflexive defender protecting the first asset (with defender value 10) would suffer an attack on the second asset, and be left with a residual value of $10 + 1 = 11$, while a defender that chose to protect the second asset (with defender value 5) would suffer an attack on the third asset, and be left with a residual value of $10 + 5 = 15$.

It follows from Property 1 that in the case of optimal attack and defense:

$$V_*(a, b) = \left( U - \sum_{i=b+1}^{b+a} v_i \right) = \sum_{i=1}^{b} v_i + \sum_{i=b+a+1}^{n} v_i \qquad (6)$$

and

$$D_*(a,b) = \left[\left(U - \sum_{i=b+1}^{b+a} v_i\right) - \left(U - \sum_{i=1}^{a} v_i\right)\right]/U$$

$$= \frac{\left[\sum_{i=1}^{a} v_i - \sum_{i=b+1}^{b+a} v_i\right]}{U} \tag{7}$$

**Property 2 (Monotonicity of Residual Value)**

Consider the case where (1) the attacker is optimizing and (2) the defender is reflexive. Then the residual system value $V(a,b)$ is decreasing with $a$ and increasing with $b$; i.e.,

$$V(a+1,b) \leq V(a,b)$$

$$V(a,b) \leq V(a,b+1)$$

**Proof:** By direct inspection of the derivations above.

**Property 3: Same/Different Valuations.** The residual value in the case of same values would be not less than the residual value in the case of differing values:

$$V_{same}(a,b) \leq V_{diff}(a,b)$$

when (1) the attacker is optimizing his value and (2) the defender is optimizing her value.

**Proof:** We denote the defender and attacker values for the surviving assets as $V(a,b)$ and $W(a,b)$, respectively. Suppose $(x^s, y^s)$ is an optimal defense strategy and attack strategy pair, in the case where the values of the assets are the same to both attacker and defender; i.e., $V(a,b) = W(a,b)$. From Prop. 1, we know that in this case, the defender protects the $b$ highest ranked targets, and the attacker strikes targets with ranks $b+1 \ldots b+a$. This is the maximal damage the attacker could inflict, as viewed by the defender.

Now, suppose the attacker's values change; i.e. $V(a,b) \neq W(a,b)$. The defender can continue with defense $x^s$. If the attacker maintains the same attack strategy $y^s$, the residual value will be unchanged, and if he changes the attack strategy, the residual value $V(a,b)$ would either increase or remain the same; it cannot decrease, because $y^s$ minimizes the residual value. The defender could even find an improved strategy $x^d$, which would further increase the residual value. QED.

**Property 4 (Monotonicity of Defensibility)**: Defensibility is monotonically increasing in $a$ and $b$, when (1) the attacker is optimizing and (2) the defender is reflexive.

**Proof (monotonicity in $a$):**

$$D(a+1,b) - D(a,b) = \frac{\left[\sum_{i=1}^{a+1} v_i - \sum_{i=b+1}^{b+a+1} v_i\right] - \left[\sum_{i=1}^{a} v_i - \sum_{i=b+1}^{b+a} v_i\right]}{U}$$

$$= \frac{v_{a+1} - v_{b+a+1}}{U} \geq 0$$

where the inequality follows because the assets are sorted in order of decreasing value.

**Proof (monotonicity in $b$):** This follows directly from monotonicity of residual value (Property 2).  QED.

**Property 5: Sensitivity of defensibility to a and b**

Defensibility is more sensitive to increasing $b$ than to increasing $a$ iff $b \leq a$.

**Proof:**

$$D(a+1,b) - D(a,b) - [D(a,b+1) - D(a,b)] = D(a+1,b) - D(a,b+1)$$

$$= \left[\frac{\sum_{i=1}^{a+1} v_i - \sum_{i=b+1}^{b+1} v_i}{U}\right] - \left[\frac{\sum_{i=1}^{a} v_i - \sum_{i=b+2}^{b+a+1} v_i}{U}\right] = \left[\frac{v_{a+1} - v_{b+1}}{U}\right] \leq 0 \text{ iff } b \leq a.$$

## 5.2 Stochastic attackers

This section is concerned with attacks that cannot be predicted with certainty.  This is particularly relevant for disruptions from accidents or natural events, but may also be representative of attackers that are opportunistic, or simply not well understood by the defender.

Recall that we distinguish between:

(1) Stationary attack: the attacker (or disruption) does not change the probability of attacking a particular asset in response to defensive actions.

(2) Uniform random attack: a special case of a stationary attack that attacks all assets with equal probability.

A precise specification of the uniform random attack is as follows: The attacker selects $a$ targets out of $n$ possible targets.  Thus, the probability that a given asset will be targeted is given by $\frac{\binom{n-1}{a-1}}{\binom{n}{a}} = \frac{a}{n}$, as expected by intuition.

**Property 6: Optimality of a reflexive defense for uniform random attacker**

The reflexive defense strategy is optimal when the attacker is uniform random.

**Proof:** Absent defense, the residual value is given by

$$V_R(a, 0) = \left(1 - \frac{a}{n}\right) \sum_{i=1}^{n} v_i \qquad (8)$$

Given $n$ assets, defense budget $b$ and $a$ strikes, the increase in the residual value from defending a set $X$ is given by $\Delta = \frac{a}{n} \sum_{i \in X} v_i$. The maximal gain is achieved by defending the $b$ highest-rank assets. QED.

From here we obtain that the optimal defense to random attacks achieves a residual value given by:

$$V_R(a, b) = \sum_{i=1}^{b} v_i + \left(1 - \frac{a}{n}\right) \sum_{i=b+1}^{n} v_i \qquad (9)$$

and a defensibility of:

$$D_R(a, b) = \frac{a}{n} \sum_{i=1}^{b} v_i \Big/ U \qquad (10)$$

**Property 7: Monotonicity of Residual Value for Uniform Random Attack**

For uniform random attackers and optimal defense, the residual value is monotonically decreasing with $a$ and increasing with $b$.

**Proof**: The result is true by inspection of the functional form of $V_R(a, b)$.

**Property 8: Monotonicity of Defensibility in the Uniform Random Case**

When the attacker is uniform random, defensibility is increasing in both $a$ and $b$.

**Proof:** The result is immediate from the linear form of the function for $D_R(a, b)$.

Notice that both here and in the deterministic case, defensibility is increasing in both $a$ and $b$.

**Property 9: Defensibility against Random and Optimal Attackers**

Let $D_*(a, b)$ and $D_R(a, b)$ be the defensibility against optimizing and uniform random attackers, respectively, with an optimizing defender. Then we have

$$D_*(a, b) \geq D_R(a, b)$$

if and only if

$$\sum_{i=1}^{a} v_i \geq \sum_{i=b+1}^{b+a} v_i + \frac{a}{n} \sum_{i=1}^{b} v_i \qquad (11)$$

**Proof:** By direct expansion.

This implies that under mild conditions, defensibility is greater against optimizing attackers. For example, it is sufficient that

(1) $a \ll n$; i.e. the number of assets is large, and

(2) Any consecutive set of $a$ assets (by rank) has a sufficiently larger value than the following set of $a$ assets. As a special case, it sufficient that $v_i > v_{i+1} + \epsilon$ for all $i$ and some constant $\epsilon > 0$.

The result appears paradoxical, because it suggests that it can be easier in some cases to defend against "better" attackers. To see why this happens, consider a system with a large number of assets but only a few high-value assets. In this system, defensibility against an optimal attacker is quite high because the defender can achieve large gains by protecting her top assets, but against a uniform random attacker, defensibility is low because the attacker is unpredictable, and is likely to target assets that have not been protected.

The opposite case occurs when the system has many comparable high-value assets. It may actually be easier to defend such a system against a random attacker than against an optimizing attacker. In particular, in such a system, a random attacker may occasionally target a defended asset purely by chance (yielding a significant benefit of defense), but for modest values of $b$, there is little gain from defending against an optimizing attacker (who will just be deflected to another target that is almost as valuable as those that have been defended).

For a numerical example, consider the case where $n = 3$, $a = 2, b = 1$. We then get $D_*(2,1) = v_1 - v_3$ and $D_R(2,1) = \frac{2}{3}v_1$. Suppose the asset values are $9.0, 8.5, 6.0$ — decreasing and concave. Then, $D_R(2,1) = \frac{6}{23.5} > D_*(2,1) = \frac{3}{23.5}$. However, if the values are $9.0, 3.0, 2.0$ (decreasing but convex), then we have $D_R(2,1) = \frac{6}{14} < D_*(2,1) = \frac{7}{14}$.

We now characterize the minimum number of assets for defensibility to be greater against an optimal attacker for the case of a geometric series. For example, if $v_{i+1} = \gamma v_i$ for $i = 2,3,..,n-1$ with $\gamma \in (0,1)$, then $\sum_{i=p}^{q} v_i = \gamma^{p-1} v_1 \frac{1-\gamma^{q-p+1}}{1-\gamma}$, we obtain

$$D_*(a,b) - D_R(a,b) = \frac{\gamma^{p-1} v_1}{V} \frac{(1-\gamma^a) - \gamma^b(1-\gamma^a) - \frac{a}{n}\left(1-\gamma^b\right)}{1-\gamma} \geq 0$$

$$(1-\gamma^a)[1-\gamma^b] \geq \frac{a}{n}\left(1-\gamma^b\right) \tag{12}$$

i.e., $D_*(a,b) \geq D_R(a,b)$ if $n \geq a/(1-\gamma^a)$ and any value of $b$. For $a = 2$ and $\gamma = 0.5$, defensibility against optimal attackers is higher than against random attackers for $n \geq 3$.

**Property 10: Sensitivity of defensibility to random and optimal attacker**

Consider the sensitivity of defensibility to attacker effort, in the cases (1) optimal attacker and, (2) uniform random attackers, i.e., (1) $\Delta_a D_* = D_*(a+1, b) - D_*(a, b)$ and (2) $\Delta_a D_R = D_R(a+1, b) - D_R(a, b)$. Then

$$\Delta_a D_* > \Delta_a D_R$$

if and only if

$$\frac{v_{a+1} - v_{a+b+1}}{U} > \frac{\frac{1}{n}\sum_{i=1}^{b} v_i}{U} \tag{12}$$

**Proof**: By direct calculation.

The result implies that $\Delta_a D_* > \Delta_a D_R$ when $n$ is sufficiently large (where sufficiently depends on the values of $a$ and $b$). It is not necessary to have a large value of $n$. For example, in the case of the geometric asset values (i.e., $v_{i+1} = \gamma v_i$), one can show that $\Delta_a D_* > \Delta_a D_R$ when $n \geq [(1-\gamma)\gamma^a]^{-1}$. For example, with $\gamma = 0.9$ and $a = 2$, $n = 13$ is sufficient. Notice that this expression does not depend on $b$. It is also possible to obtain $\Delta_a D_* < \Delta_a D_R$, for example, in the cases where the difference $v_{a+1} - v_{a+b+1}$ is vanishingly small.

## 6. Discussion

Security analysis to date has been intently focused on existing notions such as vulnerability and resilience. Our analysis here is based on the observation that some at-risk systems may be relatively easy to defend, while others may be difficult to defend even with considerable analytical and resource investments. Based on this observation, we proposed a new index, defensibility. We illustrated a number of properties of the defensibility function and proved a number of basic results in the important special case of discrete assets with additive values.

Defensibility can be computed from the (expected) attack damage before and after defense. Systems may have low attack damage before defense and also low defensibility, high attack damage before defense but high defensibility, high attack damage coupled with low defensibility, or (conceivably) low attack damage with relatively high defensibility. Systems with both high attack damage and high defensibility are the best candidates for defensive investments, when considering multiple critical systems competing for the same resources. By contrast, when systems have high attack damage but poor defensibility, alternatives other than traditional defense may need to be considered (e.g., deterring attacks by threats of retaliation, or intelligence and interdiction to interrupt attacks in the planning stages).

To summarize, our main contribution is to consider defensibility as a basic characteristic of system security. We argue that risk analysts and managers would benefit from considering defensibility as part of a security assessment. Indeed, as our proofs show, its properties reflect key aspects of system security in sometimes unintuitive but informative ways.

In particular, defensibility, unlike attack damage, allows policy makers to better determine which systems should receive improvements, while avoiding wasteful investments in systems that are not amenable to meaningful defense. The concept of defensibility could in principle be applied to almost any system or scenario considered by risk analysts, and thus has important implications for risk analysis theory and practice.

## 6.1 Directions for future work

There are many possible extensions of this work. An important problem is characterizing defensibility in the case of imperfect attackers that nonetheless exhibit more intelligence than the uniform random case. For example, an attacker might observe defenses imperfectly rather than optimally, and shift attack strategy accordingly, or may observe defenses perfectly, but be equally likely to attack any undefended asset. Alternatively, an attacker may have non-uniform attack probabilities (e.g., with the likelihood of choosing a given target being proportional to its valuation), but may be unable to observe system defenses. We leave such cases for the future.

It is clear that the concept of defensibility could also be applied to much more complex defense scenarios. Unlike in our examples, the cost of defending assets varies from asset to asset. Moreover, in reality no asset is ever perfectly defended; the remaining value of an asset after attack could be stochastic, or a fraction of the original asset value. More broadly, while we focused here on the case of independent assets, in many interesting cases the assets are interdependent (Buldyrev et al., 2010; Havlin et al., 2014). Indeed, critical infrastructure is often organized in the form of a network (Ganin et al., 2016; Murray & Grubesic, 2007). In that case (which we leave largely unexplored), the value of the system is not an additive function of the asset values, but for example may be a supermodular (synergistic) function of the values of those assets that survive the attack. Additionally, while our focus has been on damage, the resilience of the system could also be improved. We hypothesize that just like there are high-damage yet highly defensible systems, a similar situation may occur for resilience – low-resilience systems could have either high or low defensibility. Going further, just as defensibility examines the changes in the system value in response to defense effort, systems could be also characterized in terms of their response to increasing attack effort (e.g., whether additional attack effort causes accelerating damage, or diminishing marginal returns).

To conclude, we introduced the concept of defensibility to assist in determining how best to improve system defenses. We showed that defensibility depends in interesting ways both on the distribution of asset values in the system, and on the nature of the threat. We argue that defensibility is an important property that could applied to a variety of defense contexts.

## 7. Acknowledgements

The work of AG was supported in part by Uptake Technologies, Inc. The work of VB was supported in part by the US Department of Homeland Security (DHS) through the National Center for Risk and Economic Analysis of Terrorism Events (CREATE) under Cooperative Agreement No. 2010-ST-061-RE0001. However, any opinions, findings, and conclusions or recommendations in this document are those of the authors and do not necessarily reflect views of the Uptake or DHS. The authors would also like to acknowledge the extensive contributions of Mr. Ziyang Lu, a former master's student at the University of Wisconsin-Madison, to the calculations and the overall accuracy of the paper.